\documentclass[5p]{elsarticle}

\usepackage{lineno,hyperref}

\usepackage{gensymb}
\usepackage{amsmath}
\usepackage{amssymb}
\usepackage{caption}
\usepackage{subcaption}

\begin{document}
\begin{frontmatter}

\title{Resistive Plate Chambers for Precise Measurement of High-Momentum Protons in Short Range Correlations at R$^3$B}
\author[FCULAddress,lipLAddress]{M. Xarepe\corref{mycorrespondingauthor}}
\cortext[mycorrespondingauthor]{Corresponding author}
\ead{mxarepe@lip.pt}
\author[TUDaAddress,GsiAddress]{T. Aumann}
\author[lipCAddress]{A. Blanco}
\author[CEAAddress]{A. Corsi}
\author[FCULAdress,lipLAddress]{D. Galaviz}
\author[ChalmersAddress]{H. T. Johansson}
\author[GsiAddress]{S. Linev}
\author[GsiAddress]{B. Löher}
\author[lipCAddress]{L. Lopes}
\author[FrankAddress]{J. Michel}
\author[GsiAddress]{V. Panin}
\author[TUDaAddress,GsiAddress]{D. Rossi}
\author[lipCAddress]{J. Saraiva}
\author[ChalmersAddress]{H. Törnqvist}
\author[GsiAddress]{M. Traxler}

\address[FCULAddress]{Faculty of Science of the University of Lisbon, Lisbon, Portugal}

\address[lipLAddress]{Laboratory of Instrumentation and Experimental Particle Physics, Lisbon, Portugal}

\address[TUDaAddress]{Darmstadt University of Technology, Department of Physics, Darmstadt, Germany}

\address[GsiAddress]{GSI Helmholtz Centre for Heavy Ion Research, Darmstadt, Germany}

\address[lipCAddress]{Laboratory of Instrumentation and Experimental Particle Physics, Coimbra, Portugal}

\address[CEAAddress]{Department of Nuclear Physics, IRFU, CEA, Université Paris-Saclay, Paris-Saclay, France}

\address[ChalmersAddress]{Chalmers University of Technology, Gothenburg, Sweden}

\address[FrankAddress]{Goethe University Frankfurt, Frankfurt, Germany}

\address{for the R$^3$B collaboration}

\begin{abstract}
The Reactions with Relativistic Radioactive Beams (R$^3$B) collaboration of the Facility for Antiproton and Ion Research (FAIR) in Darmstadt, Germany, has constructed an experimental setup to perform fundamental studies of nuclear matter, using as a probe reactions with exotic nuclei at relativistic energies. 
Among the various detection systems, one of the most recent upgrades consisted on the installation of a large area, around $2$~m$^2$, multi-gap Resistive Plate Chamber (RPC), equipped with twelve $0.3$~mm gaps and readout by $30$~mm pitch strips, exhibiting a timing precision down to $50$~ps and efficiencies above $98$\% for minimum ionizing particles in a previous characterization of the detector. 
The RPC was part of the setup of the FAIR Phase 0 experiment that focused on measuring, for the first time, nucleon-nucleon short-range correlations (SRC) inside an exotic nucleus ($^{16}$C) that took place in spring 2022. The excellent timing precision of this detector will allow the measurement of the forward emitted proton momentum with a resolution of around 1$\%$. 
In beam measurements show an RPC efficiency above $95$\% and a time precision better than $100$~ps (including the contribution of a reference scintillator and the momentum spread of the particles) for forward emitted particles. 

\end{abstract}


\end{frontmatter}


\section{Introduction}
Understanding of how the nuclear force acts on protons and neutrons inside a nuclear system is one of the hot topics in nuclear physics. Whereas at long distances, the attractive nuclear force is reasonably well modeled by ab-initio calculations based on realistic nuclear interactions and currents \cite{Carl2015}. At short distances (below 1 fm) this force becomes strongly repulsive, giving rise to the appearance of nucleon pairs with very high relative momentum (above the Fermi level), a phenomenon known as Short-Range Correlations (SRCs) \cite{Buro1977}. 

\begin{figure}[h!]
  \center
  \includegraphics[width=0.8\linewidth]{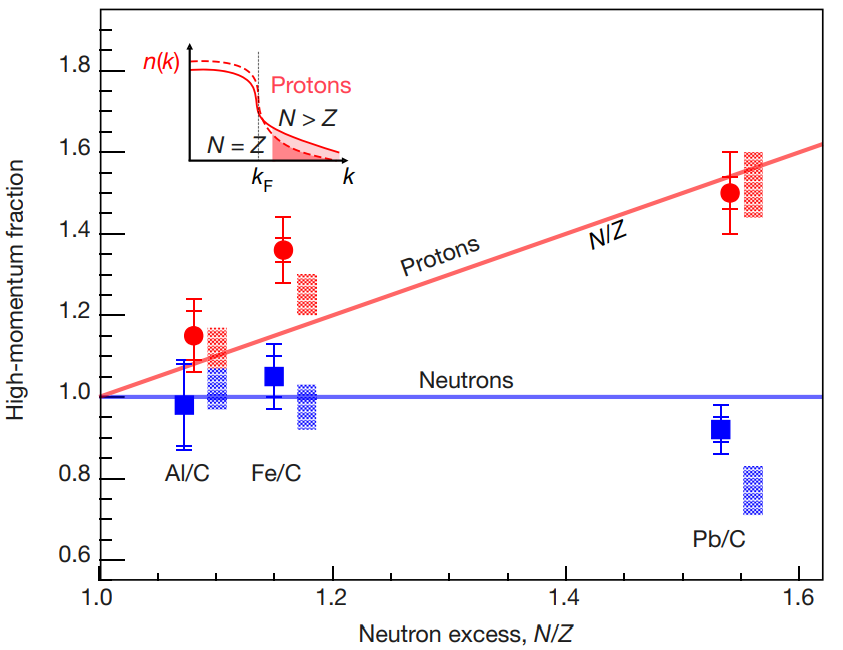}
  \caption{Relative high-momentum fraction for neutrons (in blue squares) and protons (in red circles).\cite{Clas2018}}
  \label{ex:fig:SRC}
\end{figure}

This characteristic of the strong nuclear force has been studied over the past years by the Continuous Electron Beam Accelerator Facility (CEBAF) Large Acceptance Spectrometer (CLAS) collaboration, \cite{Clas2018} in high-energy electron scattering experiments, using stable targets. One of their most relevant results indicates that the fraction of high-momentum protons increases with the neutron excess in the nucleus, see figure \ref{ex:fig:SRC}. This could have very strong implications in the description of high-density nuclear matter environments like neutron stars \cite{Souz2020}. These high-momentum nucleon pairs are directly connected to the changes observed in the quark distribution in bound nucleons inside nuclei (the so-called EMC effect) \cite{OHen2017}

As studies in stable nuclei do not allow to explore these effects at higher N/Z values, it is mandatory to orient the efforts towards exotic nuclear systems. However, reaction targets of radioactive isotopes are not available at N/Z $> 1.5$, and thus reaction studies need to be performed in inverse kinematics.

Along this line, a first SRC study using inverse kinematics was carried out in Dubna by the Baryonic Matter @ Nuclotron-based Ion Collider fAcility (NICA) (BM@N) collaboration, \cite{Patsyuk2021}. In this experiment, a stable beam of $^{12}$C, with an energy of 3.2 AGeV, collided against a proton target, studying the $^{12}$C(p,2p)X reaction. By measuring the emitted nucleons in coincidence with the residual nuclei $^{11}$B and $^{10}$Be, a limited but significant amount of SRC events were observed, paving the path towards experiments with radioactive beams using similar reaction kinematics.

As such, the Reactions with Relativistic Radioactive (R$^3$B) collaboration of the upcoming Facility for Antiproton and Ion Research (FAIR) has designed and constructed an experimental setup. This setup enables the complete characterization of the reaction products from collisions of unstable radioactive beams on a liquid hydrogen target, allowing exclusive measurement of the A(p,2pN)A-2 reaction. 

\begin{figure}[h!]
  \center
  \includegraphics[width=0.9\linewidth]{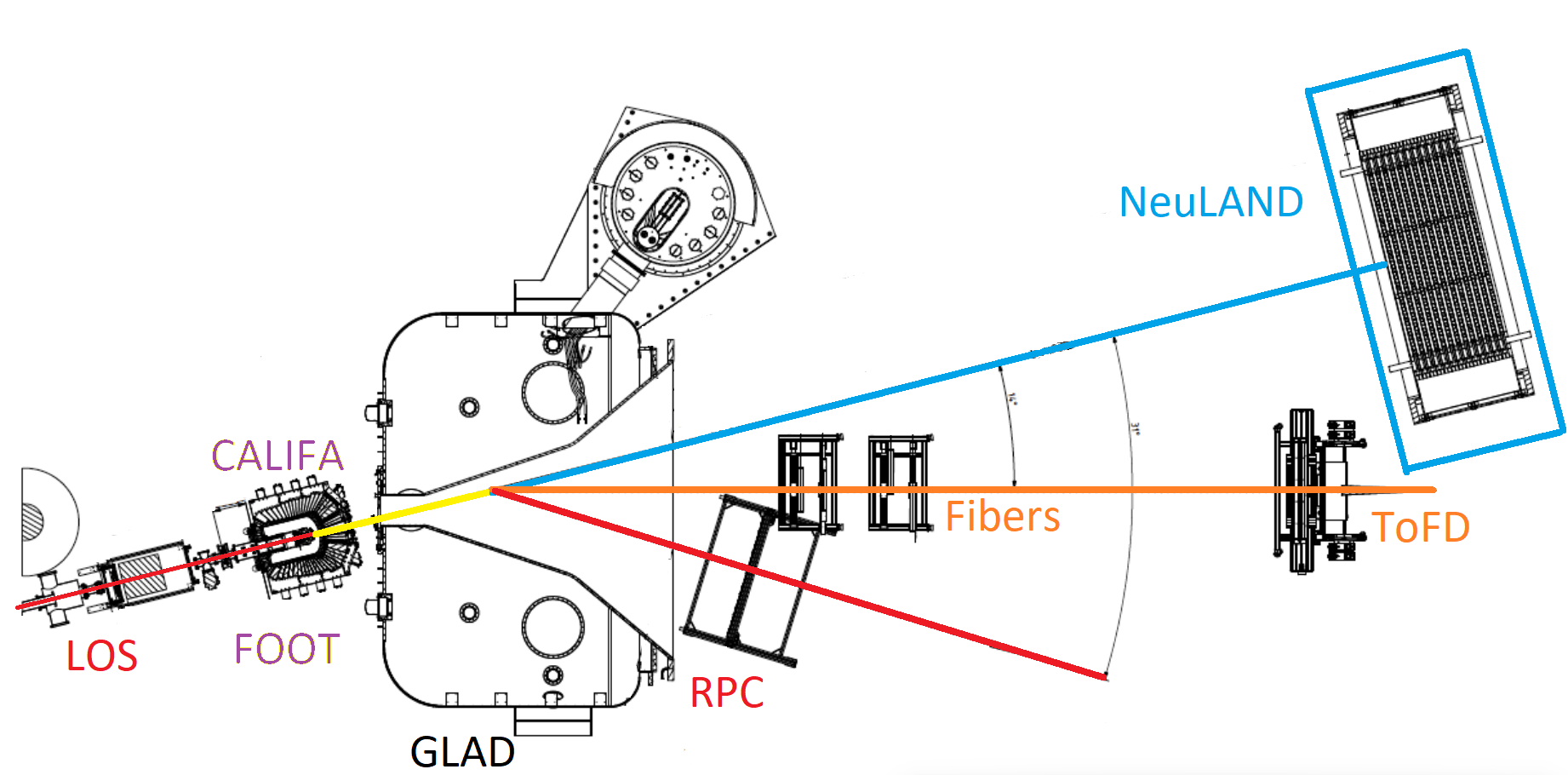}
  \caption{Schematic view of R$^3$B collaboration experimental setup considered for the measurement of SRCs in inverse kinematics.}
  \label{ex:fig:2}
\end{figure}

Figure \ref{ex:fig:2} shows the experimental setup designed by the R$^3$B collaboration for this type of measurement, from left to right in the beam direction. 

A fast scintillator (LOS) for measuring the start time of the reaction, the CaLorimeter for In-Flight detection of gamma rays and high-energy charged pArticles (CALIFA), a silicon-based detector (FOOT) for reconstructing the reaction vertex, the GSI Large Acceptance Dipole (GLAD), a superconducting magnet, a scintillator fiber tracker (Fibers), and a scintillator (TOFD) for measurement of the outgoing heavy fragment, the new Large-Area Neutron Detector (NeuLAND) for neutron TOF determination, and a Resistive Plate Chamber (RPC) for proton Time-Of-Flight (TOF) detection.

The RPC detector has been the latest to be incorporated the R$^3$B setup with the objective of precisely measuring, by means of the TOF technique, the momentum of the protons emitted in the forward direction. 

\begin{figure}[h!]
  \center
  \includegraphics[width=\linewidth]{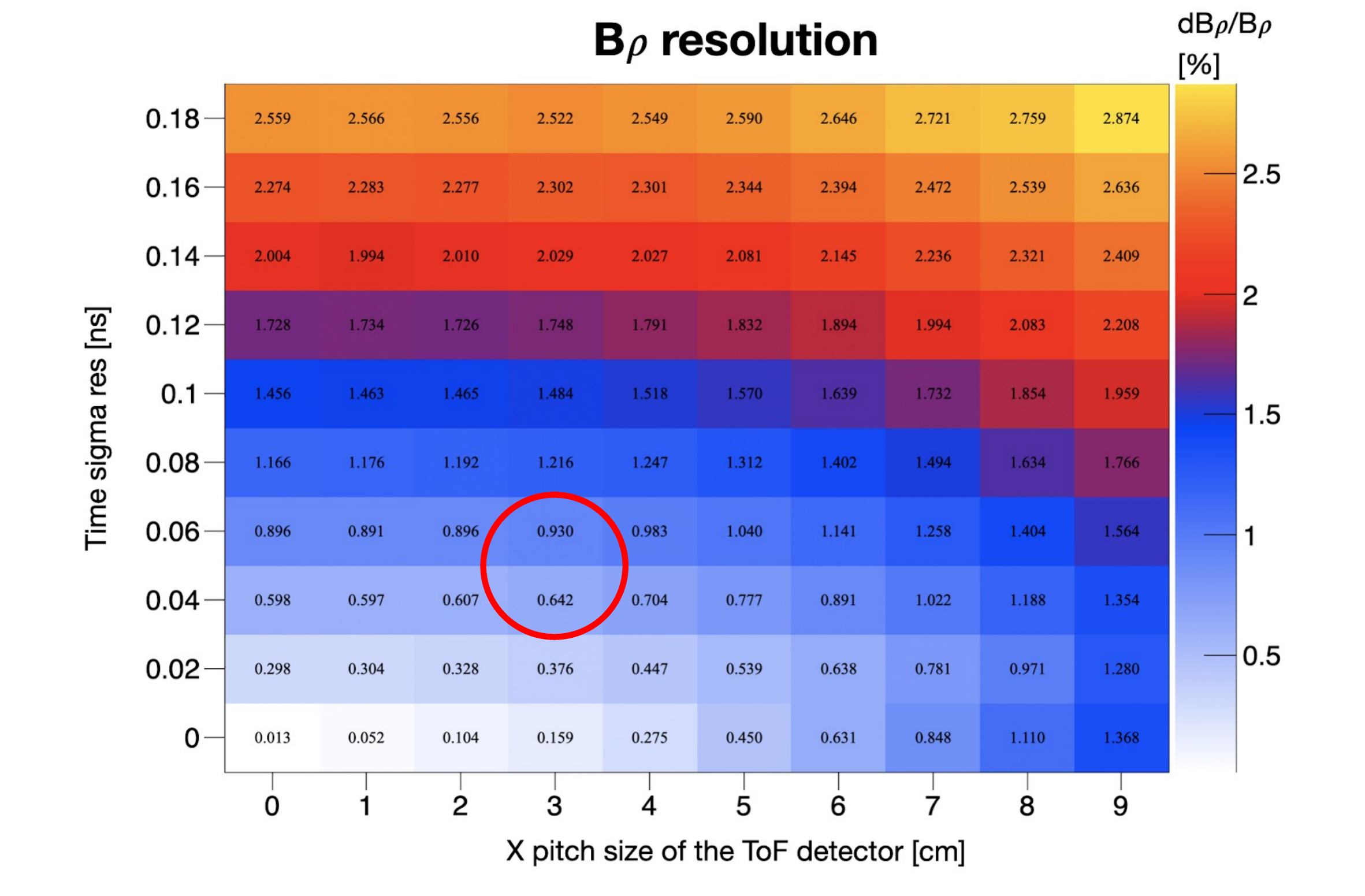}
  \caption{Resolution of the B$\rho=p/e$ reconstruction considering different detector characteristics. The RPC proposed will be located in the region delimited by the red circle, with a resolution better than 1\%.}
  \label{ex:fig:3}
\end{figure}

This detector has been previously characterised, achieving a time precision of around $50$~ps $\sigma$ and a position resolution of approximately $1$~cm$^{2}$ $\sigma$ for Minimum Ionising Particles (MIPS) \cite{Blanco2020}. A study based on simulations shows that such a detector will allow the measurement of the momentum of forward emitted protons with a resolution better than $1$\%. This result is shown in figure \ref{ex:fig:3}, where the resolution of the rigidity, B$\rho=p/e$, where $B$ is the magnetic field of GLAD, $\rho$ the radius of curvature, $p$ the proton momentum and $e$ the electron charge, is shown as a function of time resolution and pitch size. The RPC detector proposed here, located in the region delimited by the red circle.

This setup was used for the first time to measure SRC in reactions of a $^{16}$C radioactive beam, with an energy of $1.25$~AGeV, on a liquid hydrogen target in May 2022. This article concentrates on the characteristics of the RPC detector used, the calibration process, and the first results. 

\section{The RPC system on the R$^3$B setup}
\label{sec:setup}
The RPC detector consists of two multigap RPC \cite{CERRONZEBALLOS1996132} modules, each of them confined in a permanently sealed plastic\footnote{Poly(methyl methacrylate) (PMMA) for the frame and Polycarbonate (PC) for the covers.} gas tight box equipped with feed-throughs for gas and High Voltage (HV) connections. Each RPC module has six gas gaps defined by seven $1$~mm thick float glass\footnote{Bulk resistivity of $\approx 4x10^{12}$~$\Omega$cm at $25$~$^\circ$C} electrodes of about $1550~\times~1250$~mm$^2$ separated by $0.3$~mm nylon mono-filaments. The HV electrodes are made up of a semi-conductive layer\footnote{Based on an artistic acrylic paint with around $100~M\varOmega/\square$.} applied to the outer surface of the outermost glasses with airbrush techniques. 

\begin{figure}[h!]
  \center
  \includegraphics[width=\linewidth]{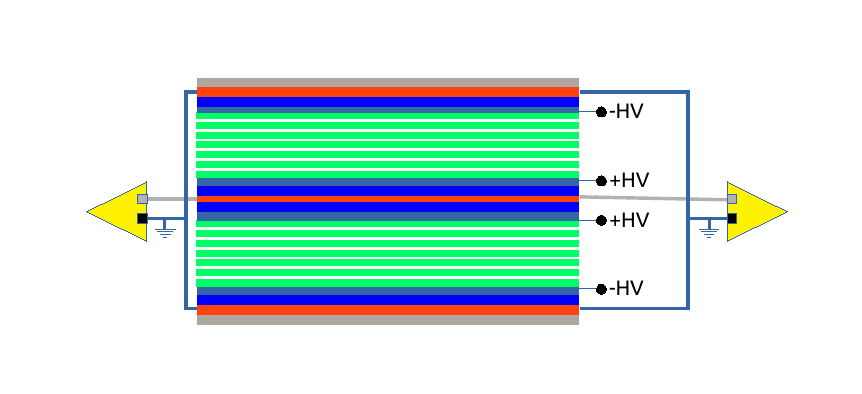}
  \caption{Schematic of the internal structure of the of RPC detector (not to scale). Grey - Aluminum box. Red - Readout electrodes. Blue - Plastic tight box. Black - HV electrodes. Green-Glass Electrodes. Yellow triangles - FEE.}
  \label{ex:fig:4}
\end{figure}

The two modules are read out in parallel by a readout strip plane\footnote{Made of $1.6$ mm Flame Retardant 4 (FR4) Printed Circuit Board (PCB).} equipped, in one side, with 41 copper strips ($29$~mm width, $30$~mm pitch, and $1600$~mm long) located in between the two modules. Two ground planes, located on top and bottom of the two-module stack, complete the readout planes. The complete structure is enclosed in an aluminum box that provides the necessary electromagnetic insulation and mechanical rigidity. In figure \ref{ex:fig:4} a schematic of the inner structure of the module is shown.

Strips are read from both sides by fast Front End Electronics (FEE) \cite{HADES_FEE}. These are capable of encoding in a single output signal: time (leading edge), with precision $<~30$~ ps $\sigma$ and charge (pulse width). The charge is obtained by measuring the Time over Threshold ($T_{\mathrm{ToT}}$) on a copy of the amplified signal, integrated with an integration constant of approximately $100$~ns. The resulting signals are read out by TDC-and-Readout Board (TRB)\cite{TRB3}, version 3, equipped with $128$ multihit Time-to-Digital Converter (TDC) channels (TDC-in-FPGA technology) with a time precision better than $20$~ps $\sigma$. This board together with a TRB, version 3sc, integrate an autonomous DAta acQuisition (DAQ) system that exports data and synchronizes, via white rabbit protocol, with the R$^3$B DAQ system. 
 
The RPC was operated in an open gas loop with a mixture of $98$\%  C$_{2}$H$_{2}$F$_{4}$ and $2$\% SF$_{6}$ at a pressure a few millibars below atmospheric pressure. In this way, the width of the gaps is correctly defined as a result of the compression exerted by the atmospheric pressure. The detector working point was measured to be about $3000$~kV/gap. 

\begin{figure}[h!]
  \center
  \includegraphics[width=0.9\linewidth]{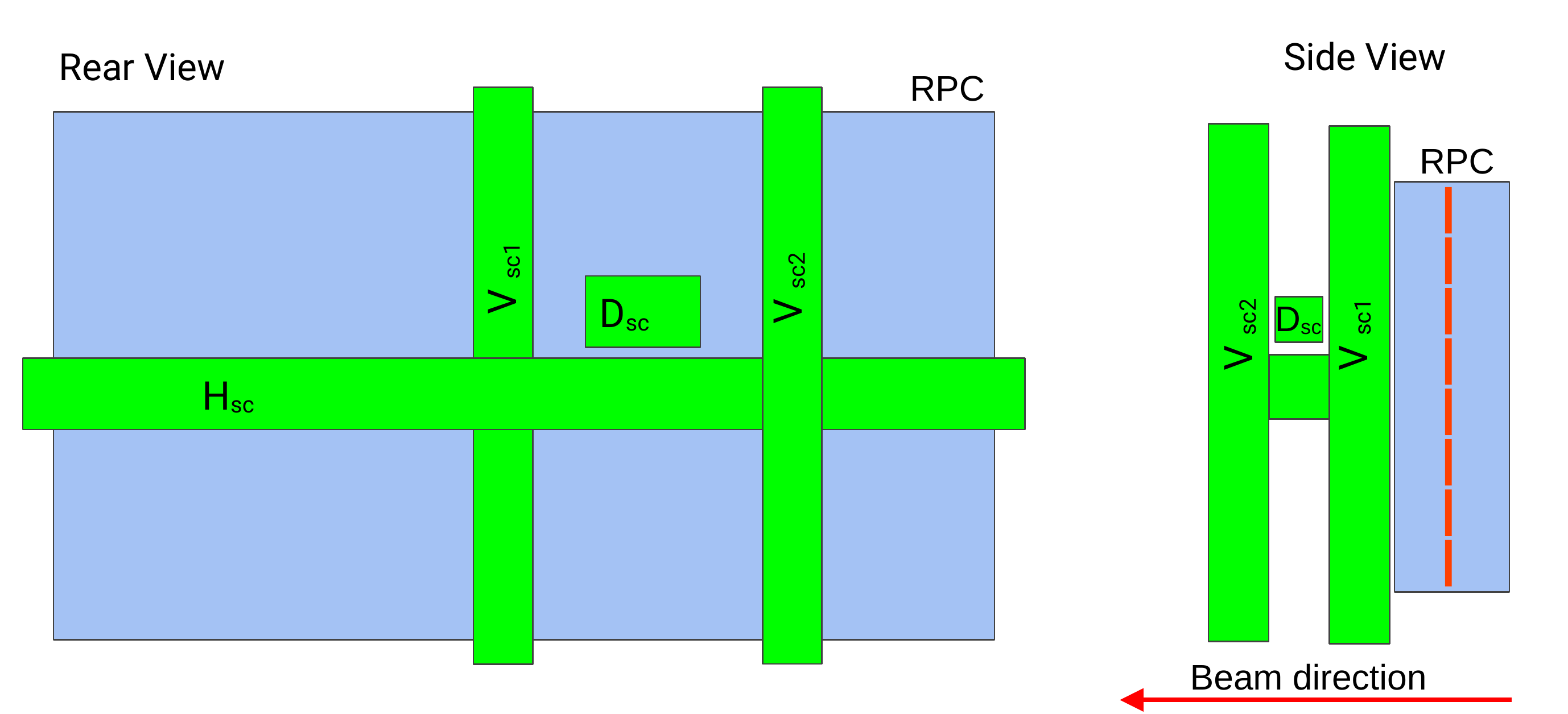}
  \caption{Schematic of the positioning of the scintillator behind the RPC, rear view on the left and side view on the right.}
  \label{ex:fig:scint}
\end{figure}

Three scintillation bars \cite{neu}, two verticals, V$_{sc1}$ and V$_{sc2}$ and one horizontal, H$_{sc}$ together with a small, $80$~mm long, scintillator, D$_{sc}$,  were placed behind and parallel to the surface of the RPC, as shown in Figure \ref{ex:fig:scint}, for calibration purposes. The scintillators are read out on both sides by photomultiplier tubes (PMT), the output signals of which are connected to the same FEE that reads the RPC but without the amplification stage. 

For each RPC strip, $i$, with signals on both sides, the $T_{\mathrm{ToT}}$ measured on the right and left sides, $T_{\mathrm{ToT},r,i}$ and $T_{\mathrm{ToT},l,i}$ will be corrected with a calculated offset associated with each of the FEE and TDC involved, $\epsilon_{\mathrm{ToT},{r,i}}$ and $\epsilon_{\mathrm{ToT},{l,i}}$. Computing in this way the charge of each channel,

\begin{equation}
\label{eqn:tot_calc}
    \begin{split}
    Q_{r,i} = T_{\mathrm{ToT},r,i} + \epsilon_{\mathrm{ToT},{r,i}}\\
    Q_{l,i} = T_{\mathrm{ToT},l,i} + \epsilon_{\mathrm{ToT},{l,i}}
    \end{split}
\end{equation}

Furthermore, the strip, $I$, with maximum $Q_{r,i}$ and $Q_{l,i}$, will be the assigned strip of the event and the charge, $Q$, can be calculated from $Q_{r,I}$ and $Q_{l,I}$ as follows.

 \begin{equation}
\label{eqn:tot_calc_2}
    Q = \frac{Q_{r,I} + Q_{l,I}}{2}
\end{equation}

Having assigned a strip to the event, the transversal position (across the strips), $Y$, can be computed using the number of the strip $I$ and the pitch of the strip, $w$,

\begin{equation} 
\label{eqn:pos_calcY}
    Y = (I-1) \times w
\end{equation}

For both the time $T$ and the longitudinal position (along the strip) $X$, the measured times on the right and left sides of the selected strip will be used, $T_{r,I}$ and $T_{l,I}$ respectively, being that for each of these variables an offset must be calculated, $\epsilon_{T,I}$ and $\epsilon_{X,I}$.

To calculate the longitudinal position, $X$, equation \ref{eqn:pos_calcX} will be used, where $V_{\mathrm{strip}}$ is the propagation velocity of the signals on the strips, $165.7$~mm/ns.

\begin{equation}
\label{eqn:pos_calcX}
    X = (\frac{(T_{r,I} - T_{l,I})}{2} \times V_{\mathrm{strip}}) + \epsilon_{X,I}
\end{equation}

For time $T$ the equation \ref{eqn:time_calc} will be used.

 \begin{equation}
\label{eqn:time_calc}
    T = \frac{(T_{r,I} + T_{l,I})}{2} + \epsilon_{T,I}
\end{equation}

\section{Calibration}
\label{sec:calibration}

The calibration of the RPC was divided into several steps.

\subsection{Time over Threshold calibration}
\begin{figure}[h!]
  \center
  \includegraphics[width=\linewidth]{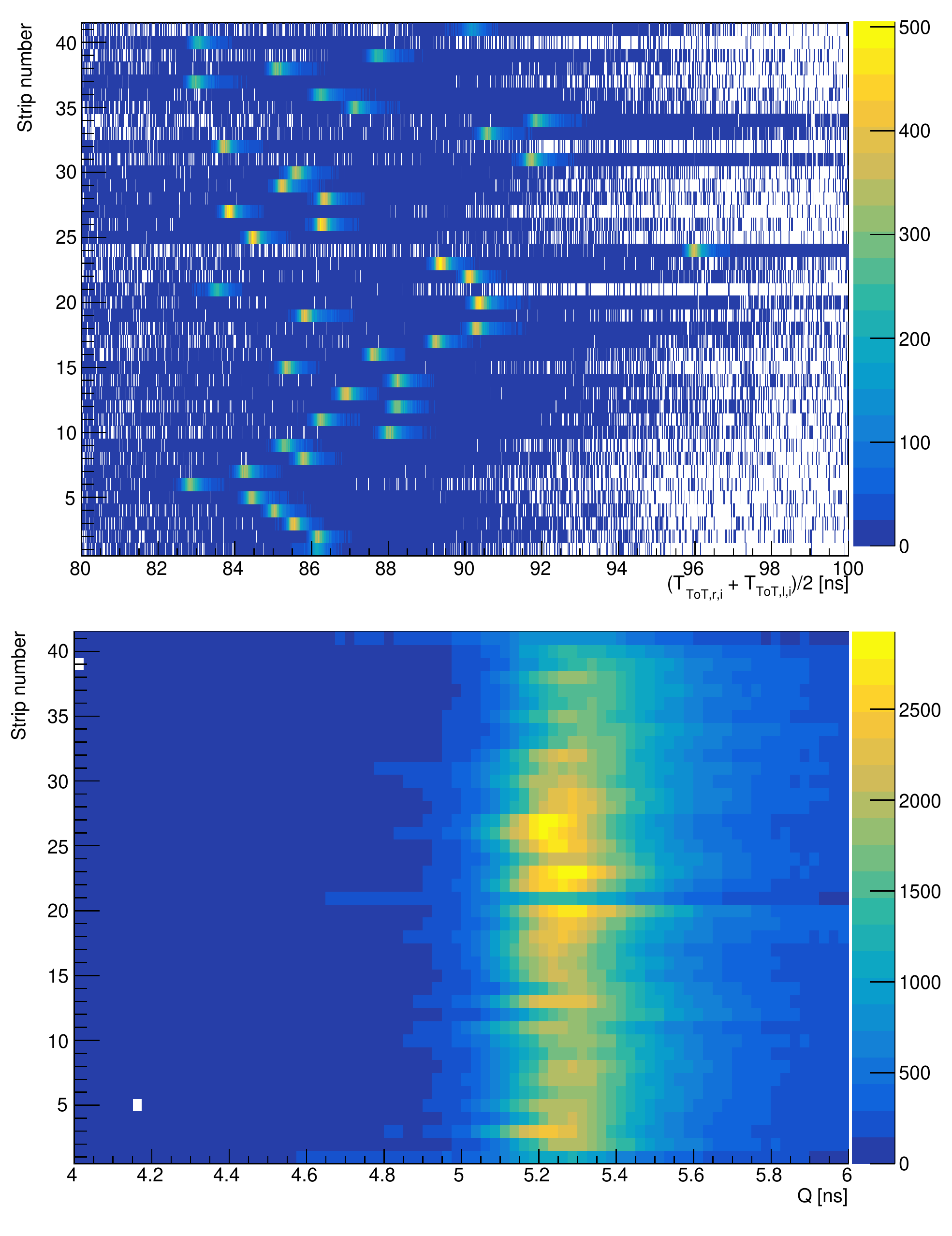}
  \caption{Top,  $(T_{\mathrm{ToT},r,i}+T_{\mathrm{ToT},l,i})/2$ and bottom, $Q$ for all strips.}
  \label{ex:fig:5}
\end{figure}
The calibration of the $T_{\mathrm{ToT}}$ consists in the calculation of the $\epsilon_{\mathrm{ToT,}{r,i}}$ and $\epsilon_{\mathrm{ToT,}{l,i}}$ parameters, originated by the different response of each of the FEE and TDC channels involved, which align the minimum value of $T_{\mathrm{ToT},r,i}$ and $T_{\mathrm{ToT},l,i}$ between all channels. This calibration is crucial for the determination of $Q_{r,I}$ and $Q_{l,I}$ with the consequent correct calculation of $Q$, $T$, $Y$ and $X$.

If this calibration is not performed correctly, $Q_{r,I}$ and $Q_{l,I}$ could be attributed to different channels on the left and right sides, in which case the event would be discarded or attributed to the wrong strip. This is visible in the figure \ref{ex:fig:5} top, where the uncalibrated $(T_{\mathrm{ToT},r,i}+T_{\mathrm{ToT},l,i})/2$ spectra for each of the strips are plotted. Due to the lack of calibration, $10$\% of the events will be discarded. After a correct calibration, figure \ref{ex:fig:5} bottom, the value of events with incorrect $Q$ decreases to $2$\%. More work is being done to further improve this value. 

\subsection{Longitudinal strip position calibration}
\begin{figure}[h!]
  \center
  \includegraphics[width=\linewidth]{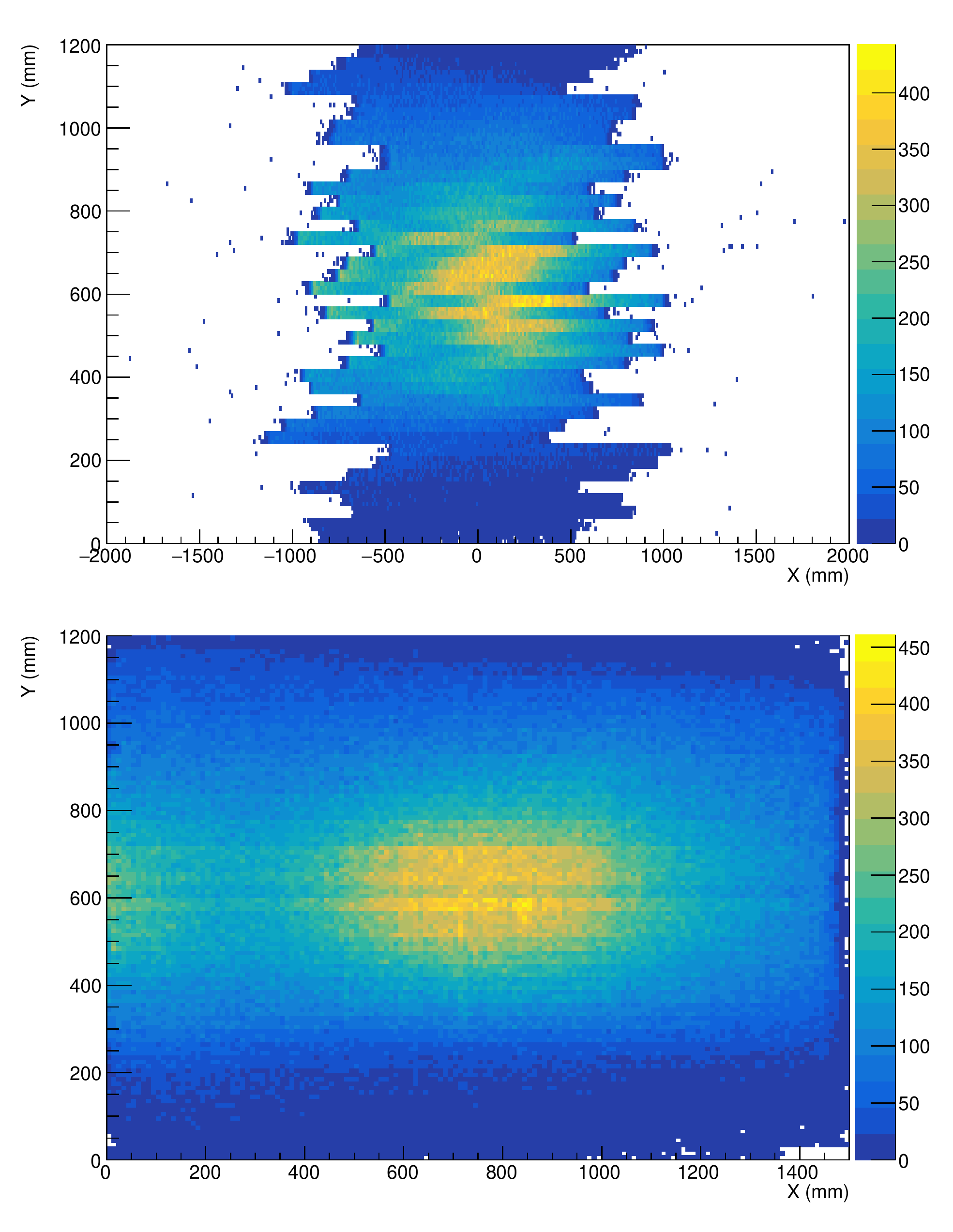}
  \caption{Top, $X$, $Y$ map without calculation of $\epsilon_{X,I}$ parameters and with calculated ones in bottom. The central spot corresponding to the forward emitted particles (protons) and the geometrical acceptance defined by the GLAD magnet (halo at the periphery of the plot) are clearly visible in bottom plot.}
  \label{ex:fig:6}
\end{figure}
After a correct calibration of the $T_{\mathrm{ToT}}$ and calculation of $Q$, the transverse and longitudinal positions, $Y$ and $X$, are calculated. For the latter, it is necessary to determine the parameters $\epsilon_{X,I}$, which have as origin the different time offsets associated with each strip, FEE and TDC channels.

This is visible in the figure \ref{ex:fig:6} top, where a two-dimensional histogram of $Y$, $X$, without the calibration parameters, is shown. In contrast, figure \ref{ex:fig:6} bottom shows the same histogram with the calculated calibration parameters, where the central spot corresponding to the forward emitted particles (protons) and the geometrical acceptance defined by the GLAD magnet (halo at the periphery of the plot) are clearly visible. 

\subsection{Strip time calibration}
\begin{figure}[h!]
  \center
  \includegraphics[width=0.9\linewidth]{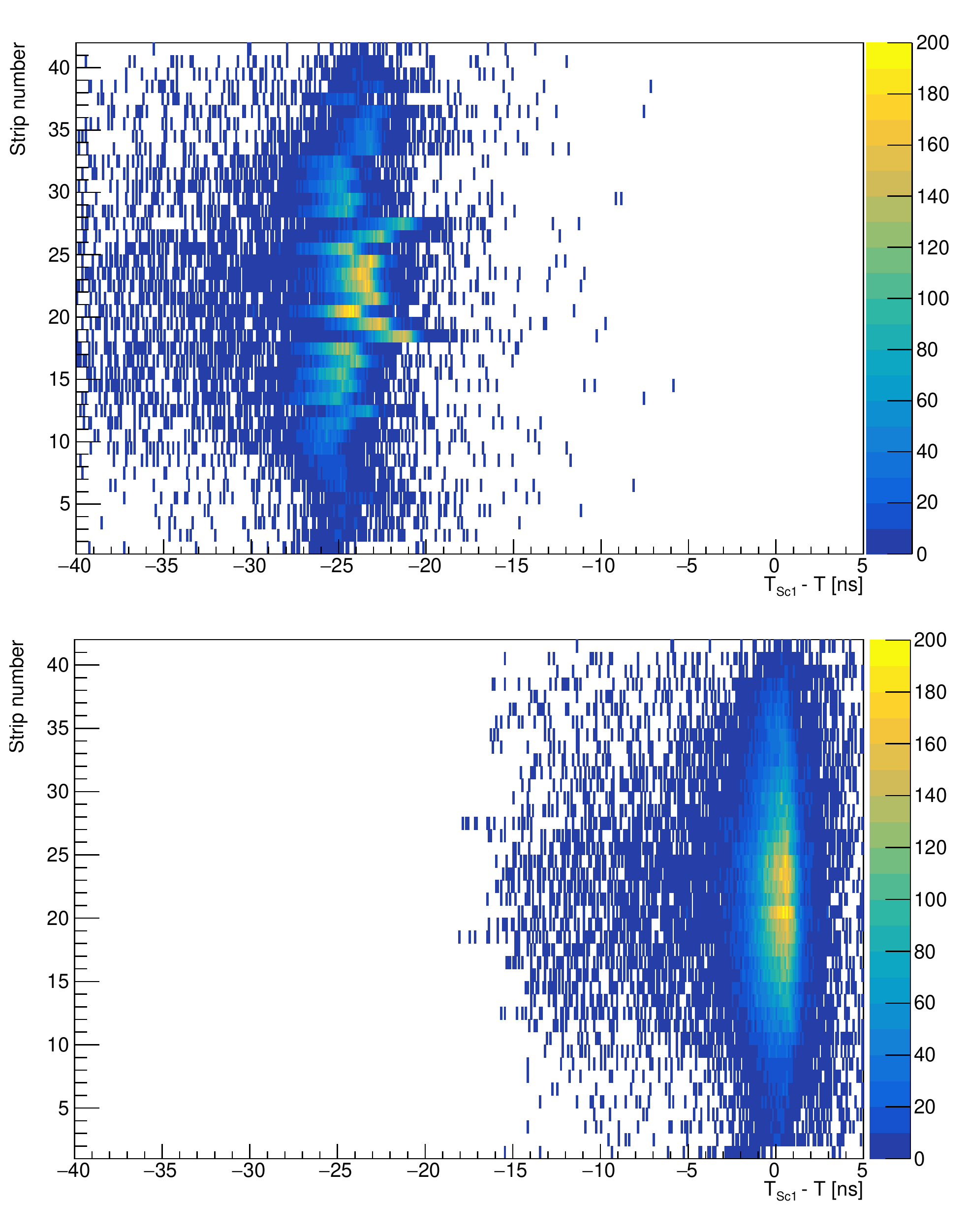}
  \caption{Top, $T_{sc1} - T$ difference without calculation of $\epsilon_{T,I}$ parameters and with calculated ones in bottom.}
  \label{ex:fig:calibrated_time}
\end{figure}

Finally, the time $T$, can be determined, for which the parameters $\epsilon_{T,I}$ must be calculated, which have as origin the different time offsets associated with each strip, FEE and TDC channels.

The vertical scintillators are used to calculate these parameters. The differences between the scintillator time and the RPC time, $T_{sc1} - T$ (for the case of the scintillator V$_{sc1}$), are calculated for events passing through each of the RPC strips. Assuming that the scintillators are perfectly parallel to the surface of the RPC, the time of flight for the particles between the two detectors should be, on average, equal, without depending on the strip of the RPC that has produced the signal. Therefore, the parameters $\epsilon_{T,I}$ are calculated such that the average of $T_{sc1} - T$ is always the same.

The result can be seen in figure \ref{ex:fig:calibrated_time}, where $T_{sc1} - T$, is calculated for each of the strips . In the top part without the calibration parameters and at the bottom with the calculated calibration parameters applied. This procedure ensures that the RPC surface always produces the same time regardless of the strip producing the signal. 

\section{Beam time performance}
The RPC autonomous DAQ system ran without interruption for more than two weeks, delivering data and correctly synchronising with the R$^3$B DAQ. 

\begin{figure}[h!]
  \center
  \includegraphics[width=0.9\linewidth]{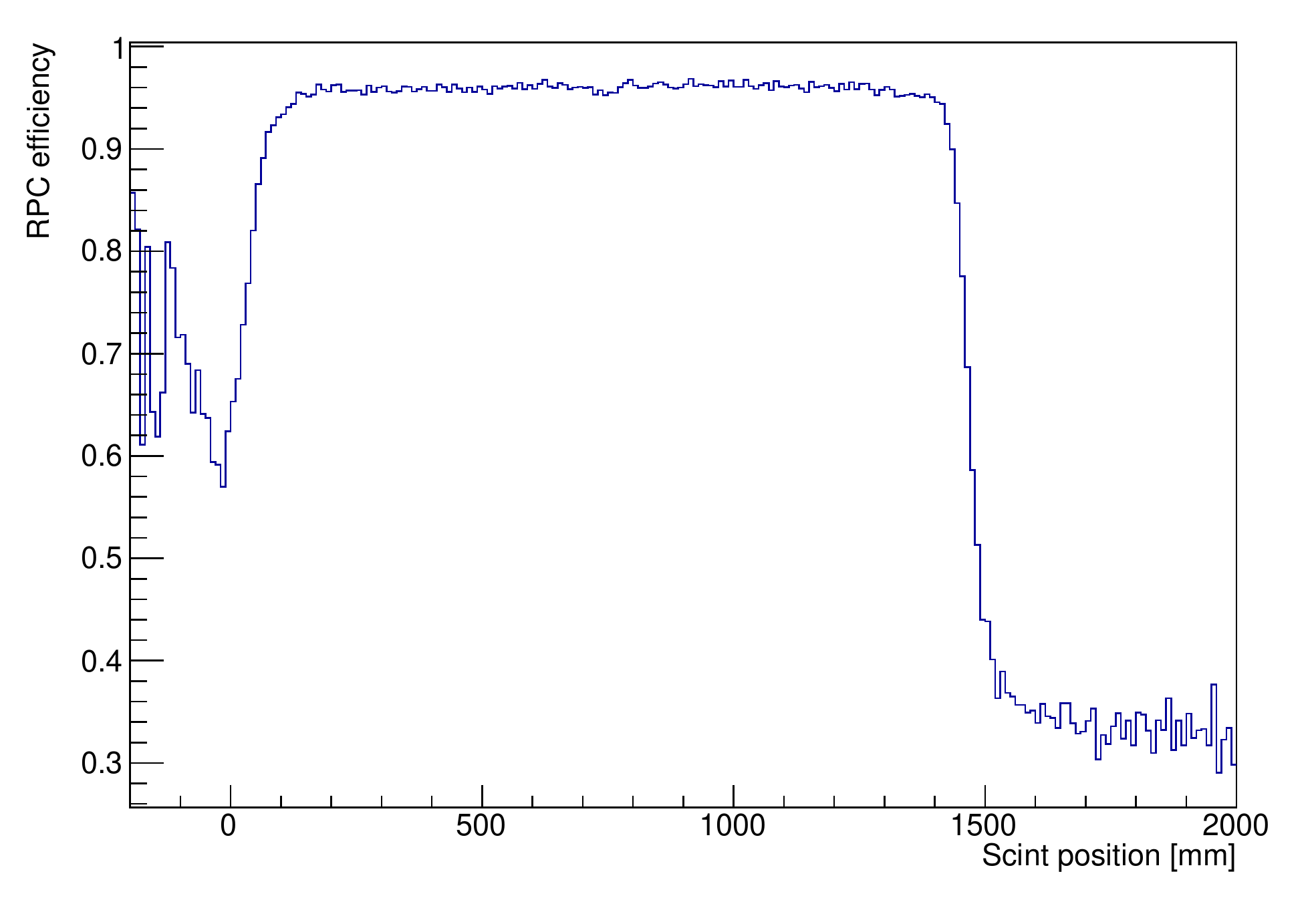}
  \caption{Beam time efficiency. Estimated efficiency of the RPC as a function of the H$_{sc}$ longitudinal position placed behind the RPC, showing an average value of around $95$\%.}
  \label{ex:fig:effi}
\end{figure}

Using the horizontal scintillator, H$_{sc}$, the efficiency was calculated on the overlapped area. It was estimated by dividing the events seen by the LOS, RPC and H$_{sc}$ by all the events seen by the LOS and H$_{sc}$ detectors. The result can be seen in figure \ref{ex:fig:effi} where the efficiency of the RPC is shown as a function of the position given by the scintillator, showing a mean value of $95$\%.

\begin{figure}[h!]
  \center
  \includegraphics[width=0.9\linewidth]{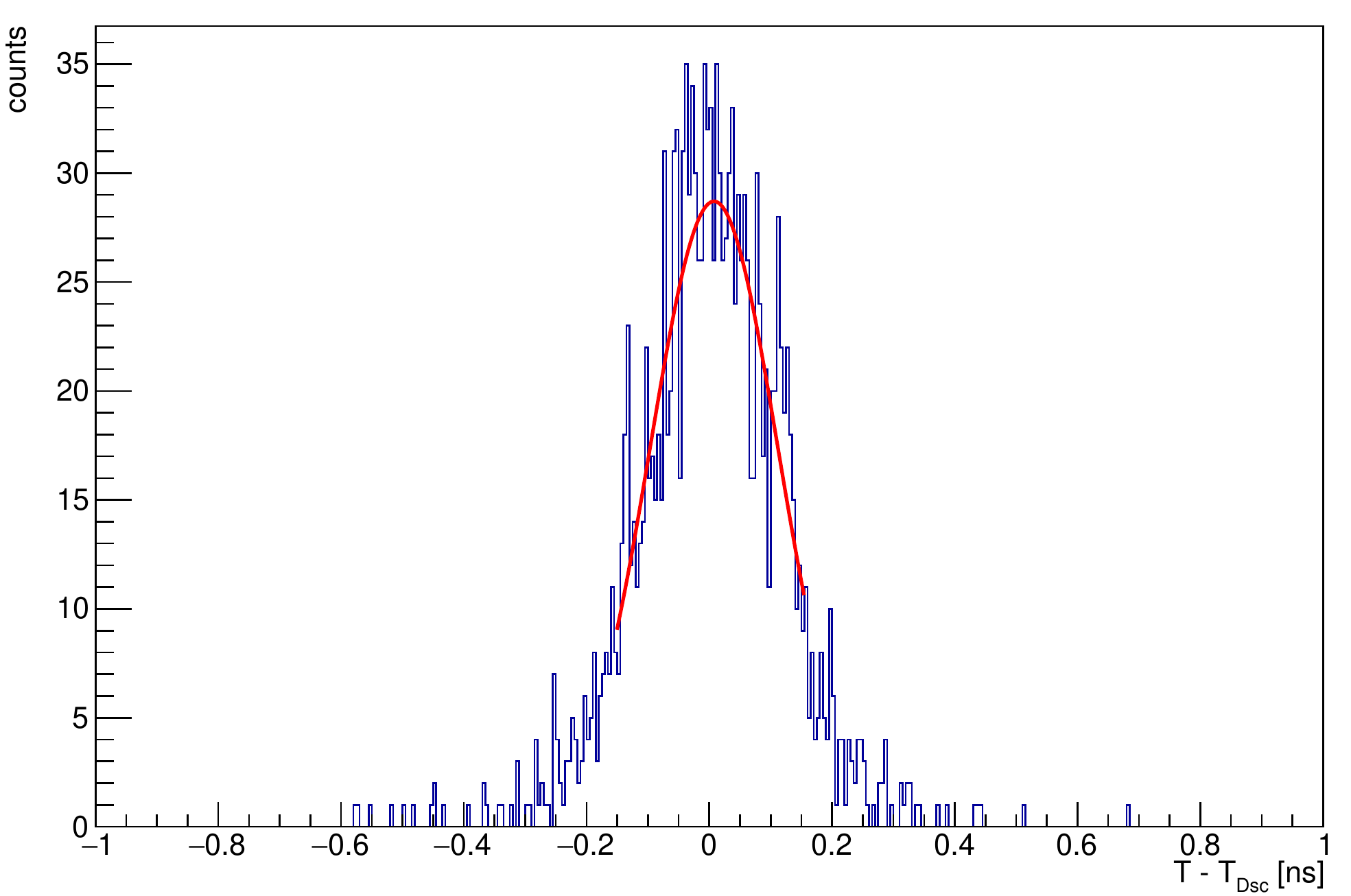}
  \caption{Time difference between the RPC and the small scintillator (D$_{sc}$) showing a sigma, after fitting the central Gaussian part of the distribution, of $100$~ps.}
  \label{ex:fig:time_diff}
\end{figure}

The time precision of the RPC was estimated by performing the time difference between the RPC and the small scintillator (D$_{sc}$), which has the best time precision of the four scintillators. Figure \ref{ex:fig:time_diff} shows the time difference $T - T_{Dsc}$, which after fitting the central Gaussian part of the distribution shows a value of $100$~ps $\sigma$. This value includes the precision of the RPC, D$_{sc}$, and the spread in momentum of the particles. The precision of D$_{sc}$ was evaluated, in a previous characterization, to be around $35$~ps $\sigma$ for a collimated monochromatic relativistic pion beam, conditions which are not met here, being its contribution probably higher. The contribution of momentum has been calculated to be negligible for the distance between RPC and D$_{sc}$ based on the measured TOF between LOS and RPC.

\section{Conclusions}
The setup (installation, DAQ integration and calibration) of a large area, $2$~m$^2$, RPC detector for the precise measurement of the momentum of the forward emitted protons in collisions of unstable radioactive beams on a liquid hydrogen target for the measurement of nucleon-nucleon short-range correlations in the R$^3$B experiment has been successfully done. The RPC exhibits during the first beam time an efficiency higher than $95$\% and a time precision better than $100$~ps $\sigma$.  

\section{Acknowledgments}
This work was supported by Fundação para a Ciência e Tecnologia (FCT), Portugal, within the framework of the project EXPL/FIS-NUC/0364/2021 and the PhD grant 2021.05736.BD.

\end{document}